\documentclass[12pt,tightenlines]{revtex4}
\usepackage{amsfonts}
\usepackage{amsmath}
\usepackage{txfonts}
\usepackage{graphicx}
\usepackage{amssymb}
\usepackage{color}
\usepackage{wrapfig,lipsum,booktabs}
\usepackage{floatrow,float}
\usepackage{epstopdf}
\usepackage{hyperref}
\usepackage[normalem]{ulem}

\newcommand{\beginsupplement}{
        \setcounter{table}{0}
        \renewcommand{\thetable}{S\arabic{table}}
        \setcounter{figure}{0}
        \renewcommand{\thefigure}{S\arabic{figure}}
     }
     
\begin{document}
\title{Topological Schemas of Memory Spaces}
\author{Andrey Babichev and Yuri Dabaghian\textsuperscript{*},}
\affiliation{Department of Computational and Applied Mathematics, Rice University, Houston, TX 77005, USA \\
$^{*}$e-mail: dabaghian@gmail.com}
\date{\today}

\begin{abstract} 
\vspace{10 mm}

\textbf{Abstract}. Hippocampal cognitive map---a neuronal representation of the spatial environment---is broadly 
discussed in the computational neuroscience literature for decades. More recent studies point out that hippocampus 
plays a major role in producing yet another cognitive framework that incorporates not only spatial, but also nonspatial 
memories---the memory space. However, unlike cognitive maps, memory spaces have been barely studied from a 
theoretical perspective. Here we propose an approach for modeling hippocampal memory spaces as an epiphenomenon 
of neuronal spiking activity. First, we suggest that the memory space may be viewed as a finite topological space---a 
hypothesis that allows treating both spatial and nonspatial aspects of hippocampal function on equal footing. We then 
model the topological properties of the memory space to demonstrate that this concept naturally incorporates the notion 
of a cognitive map. Lastly, we suggest a formal description of the memory consolidation process and point out a connection 
between the proposed model of the memory spaces to the so-called Morris' schemas, which emerge as the most compact 
representation of the memory structure.

\vspace{10 mm}

\end{abstract}

\maketitle

\newpage

\section{Introduction}
\label{section:intro}

In the neurophysiological literature, the functions of mammalian hippocampus are usually discussed from the 
following two main perspectives. One group of studies addresses the role of the hippocampus in representing the 
ambient space in a cognitive map \citep{Tolman,Moser}, and the other focuses on its role in processing nonspatial 
memories, notably the episodic memory frameworks \citep{Eichenbaum,Crystal,Dere, Hassabis}. Active studies of 
the former began with the discovery of the ``place cells''---hippocampal neurons that fire action potentials in discrete 
regions of the environment---their respective ``place fields''. It was demonstrated, e.g., that place cell firing can be 
used to reconstruct the animal's trajectory on moment by moment basis \citep{Guger,Jensen,Barbieri}, or to describe 
its past navigational experiences \citep{Carr} and even its future planned routs \citep{Dragoi}, which suggests that 
the cognitive map encoded by the hippocampal network provides a foundation of the animal's spatial memory and 
spatial awareness \citep{OKeefe,Best}.

\begin{figure} 
\includegraphics[scale=0.88]{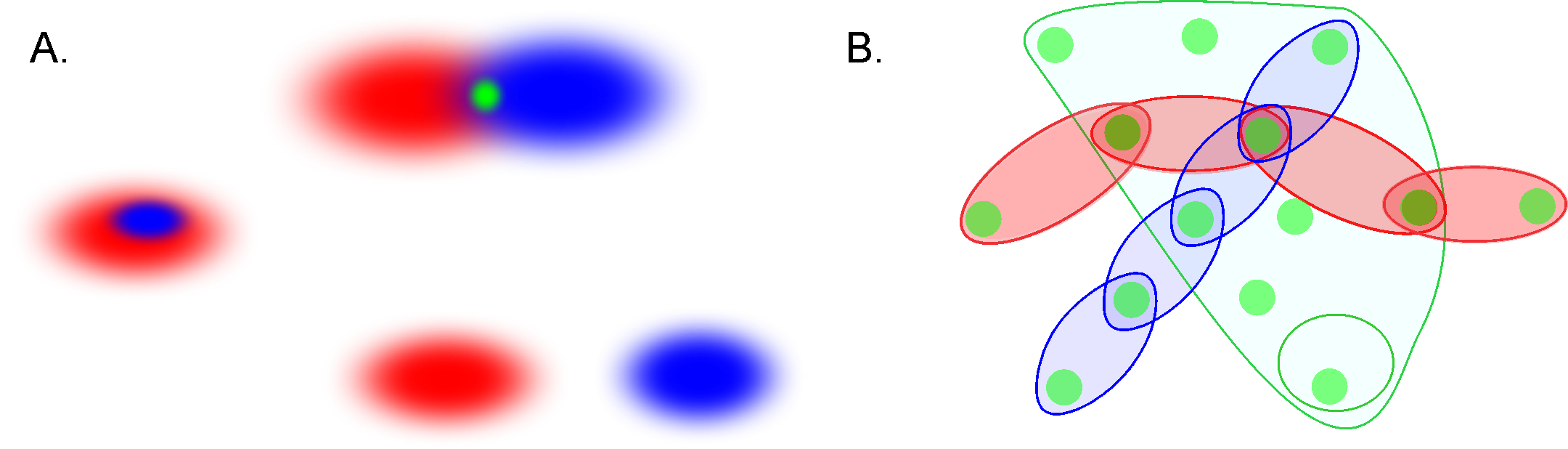}
\caption{{\footnotesize \textbf{A schematic illustration of memory space concept}. ({\bf A}) Memory elements are viewed 
as regions in memory space, $r_1$ and $r_2$ (red and blue ovals). The overlapping regions yield a smaller region in the 
intersection that represents a shared memory (top figure). Alternatively, one memory region can also contain another (the 
middle figure), or two memory regions can be separate from one another (bottom figure). 
({\bf B}). Memory elements jointly form a cohesive framework---the memory space---into which different memory sequences 
are embedded. The episodes connected in sequences can be viewed as chains of interconnected regions that run across 
the memory space, whereas memories that are ``broader in the features'' are represented by extended, space-like domains 
of the memory space. The most elementary, indecomposable elements shared between distinct behavioral episodes represent 
``nodes''---the elementary locations in the memory space.}}
\label{Fig1}
\end{figure}

On the other hand, it was observed that hippocampal lesions result in severe disparity in episodic memory function, 
i.e., the ability to produce a specific memory episode and to place it into a context of preceding and succeeding events. 
In healthy animals, episodic sequences consistently interleave with one another, yielding an integrated, cohesive semantic 
structure \citep{Agster,Fortin1, Fortin2,Wallenstein,MacDonald}. In \citep{Eichenbaum1,Eichenbaum2, Eichenbaum3} it was 
therefore suggested that the overall memory framework should be viewed as an abstract ``memory space'' $\mathcal{M}$, 
in which individual memories correspond to broadly understood ``locations'' or ``regions.'' The relationships between 
memories are represented via spatial relationships between these regions, such as adjacency, overlap or containment 
(Fig.~\ref{Fig1}). It was also suggested that the animals can ``conceptually navigate'' the memory space by perusing 
through learned associations, i.e., by comparing and contrasting directly connected memories and inferring relationships 
between indirectly linked ones \citep{Buffalo,BuzMos}. In this approach, the conventional spatial inferences that enable 
spatial navigation of physical environments based on cognitive maps are viewed as particular examples of a navigating 
a memory space, which in general allow inferring associations and producing reasoning chains of abstract nature \citep{Eichenbaum1}. 
In other words, the concept of memory space generalizes the notion of cognitive map: the latter unifies specifically spatial 
memories and hence forms a substructure or a subspace embedded into a larger memory space.

\textbf{Extended topological hypothesis}. Traditionally, the cognitive map was viewed as a Cartesian map of animal's 
locations, distances to landmarks, angles between spatial cues and so forth \citep{OKeefe,Best}. However, increasing 
amount of experimental evidence suggests that this map is based on representing qualitative spatial relationships rather 
than precise spatial metrics. For example, it has been demonstrated that if the environment gradually changes its shape 
in a way that preserves relative order of spatial cues, then the temporal order of the place cell spiking and the relative 
arrangement of the place fields remain invariant throughout the change \citep{Muller,Gothard,Lever,Leutgeb,Wills, Colgin1,Diba,Wu}. 
This suggests that place cell coactivities emphasize contiguities between locations as well as the temporal sequence 
in which they are experienced, and hence that the hippocampus encodes a flexible framework of spatial relationships---a 
topological map of space \citep{Poucet,Wallenstein,Alvernhe1,Dabaghian}.

The mathematical nature of memory space has not been addressed in computational neuroscience literature. However, 
general properties of the episodic memory frameworks suggest that such a space should also be viewed as primarily 
topological. Indeed, the ``regions'' or ``locations'' in $\mathcal{M}$ are abstract concepts that are not attributed any 
particular geometric features, such as shape or size, and the relationships between these regions do not involve precise 
metric calculations of distances and angles. Rather, the memory space is based on qualitative spatiotemporal relationships, 
which is a defining property of topological spaces \citep{Vickers}. Thus, the topological perspective provides a common 
ground for both ``spatial'' and ``non-spatial'' aspects of the hippocampal functions. In fact, the contraposition between 
these two specialties of the hippocampus might have originated, in the first place, from an excessive ``geometrization'' 
of the cognitive map. If the hippocampal spatial map is Cartesian, then it is not entirely clear which mechanism could be 
responsible for representing coordinates, distances, angles, etc., in the spatial domain and only qualitative relationships 
between memory items in the mnemonic domain. On the other hand, it is hard to attribute geometric characteristics to the 
elements of the memory space, especially to the nonspatial memories, and it is unclear what role geometry would play in 
that space. However, if both the cognitive map and the memory space are viewed as topological, based on relational 
representation of information, then the principles of spatial representation and mnemonic memory functions converge 
\citep{Dabaghian}. Taken together, these arguments suggest that the hippocampal network encodes a generic topological 
framework, which may be manifested as a cognitive map or as a more general memory space, depending on the context 
and the nature of the encoded information.

In the following, we propose a theoretical framework that incorporates both the cognitive maps and the memory spaces 
in a single model and allows interpreting hippocampal memory spaces as epiphenomena of neuronal activity. In particular, 
it allows relating the topological properties of the memory space to the parameters of the place cell spiking, e.g., spiking 
rate, spatial selectivity of firing, etc., and connecting the concept of memory space to the Morris' memory schemas.

\section{The Model}
\label{section:model}

In \citep{Babichev1} we proposed theoretical approach for modeling cognitive maps, which allows combining the 
information provided by the individual place cells into a large-scale topological representation of the environment. 
Following the standard neurophysiological paradigm, the model assumes, firstly, that the activity of each individual 
place cell $c_k$ encodes a spatial region $r_k$ that serves as a building block of the cognitive map. Secondly, it 
assumes that the large-scale structure of the cognitive map emerges from the connections between these regions,
encoded in a population place cell assemblies---functionally interconnected groups that synaptically drive their 
respective reader-classifier (readout) neurons in the downstream networks \citep{Harris,Buzsaki1}. A particular 
readout neuron integrates the presynaptic inputs and produces a series of spikes, thus actualizing a specific relationship 
$\rho(r_1,r_2,...,r_m)$ between the regions $r_1$, $r_1$,... $r_m$. 

\begin{figure} 
\includegraphics[scale=0.88]{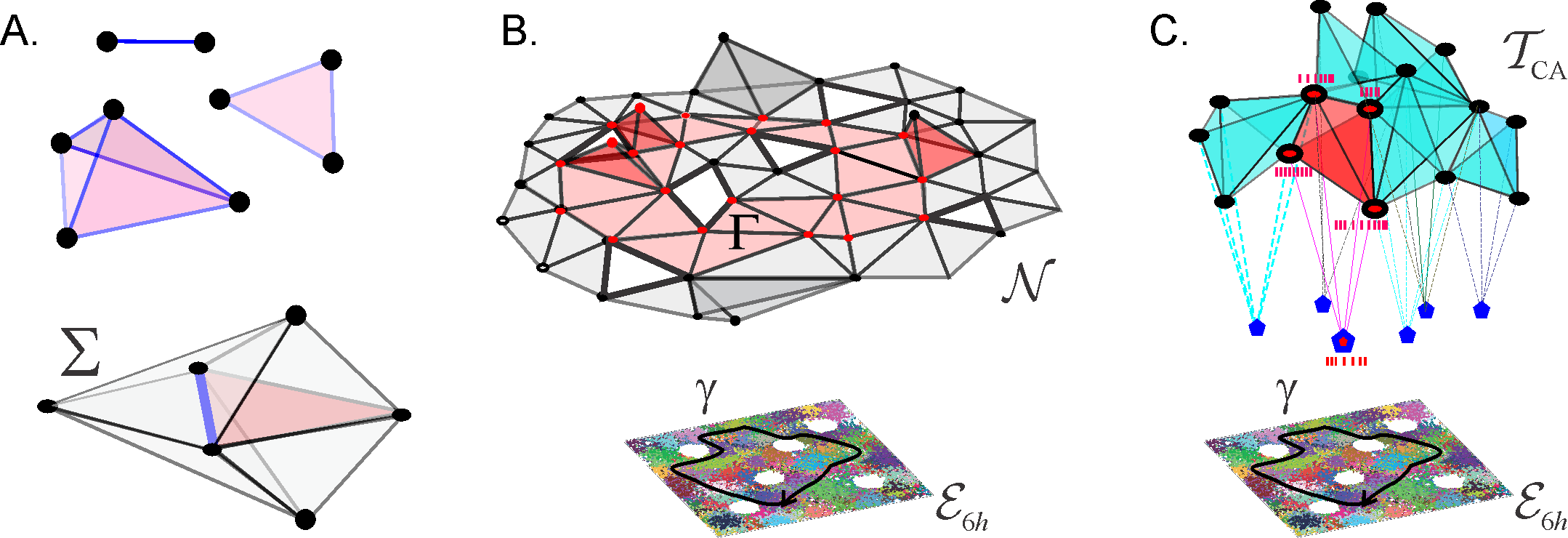}
\caption{{\footnotesize \textbf{Coactivity complex and the cell assembly complex.}: ({\bf A}). Three exemplary 
simplexes: a one-dimensional ($1D$) link, a $2D$ triangle and a $3D$ tetrahedron are shown on the top. Together 
a few simplexes form a small simplicial complex $\Sigma$ shown below. Note that the $2D$ and $3D$ simplexes 
surrounding a $1D$ simplex (the blue link) form its vicinity--this observation will be used in the Alexandrov space 
construction. ({\bf B}). The nerve complex $\mathcal{N}$ represents the pattern of overlaps between place fields 
covering a given environment, every simplex $\sigma\in\mathcal{N}$ represents a combination $\sigma = [\pi_{i_0}, 
\pi_{i_1}, ..., \pi_{i_d}]$ of overlapping place fields, $\pi_i \cap \pi_i...\cap \pi_d \neq\emptyset$. The bottom 
of the panel shows place field map, $M(\mathcal{E})$ of a square environment with six holes, $\mathcal{E}_{6h}$, 
traversed by a trajectory $\gamma$ (black line). Place cells are shown as vertices of the simplexes: the active place 
cells are shown as red points and the inactive ones as black points. 
The figure schematically represents a $2D$-skeleton of $\mathcal{T}$, used to compute the topological features of 
the underlying environment. The simplexes representing place cell combinations that become coactive as the 
animal navigates along $\gamma$ form a simplicial path $\Gamma$, shown in red. The simplicial path encircles 
the hole in the coactivity complex that represents the physical hole in the environment. 
The coactivity complex $\mathcal{T}$ is an implementation of the nerve complex in temporal domain: every simplex, 
$\sigma\in\mathcal{T}$ represents a combination of coactive place cells, $\sigma = [c_1,c_2,...,c_n]$. Over time, 
$\mathcal{T}$ becomes structurally identical to $\mathcal{N}$.
({\bf C}). Simplexes of the cell assembly complex $\mathcal{T}_{CA}$ represent the cell assemblies, shown as 
interconnected cliques of vertexes---that jointly drive readout neurons in the downstream networks (shown as 
pentagons to which place cells connect synaptically). Red clique represents an ignited place cell assembly, eliciting 
a spiking response from its readout neuron.}}
\label{Fig2}
\end{figure} 

A few schematic models were built in \citep{Dabaghian2,Arai,Basso,Hoffman,Babichev1,Babichev2} based on the 
observation that an assembly of place cells $c_1, c_2, ..., c_m$, can be schematically represented by an ``abstract 
simplex,'' $\sigma = [c_1, c_2, ..., c_m]$. In mathematics, the term ``simplex'' usually designates a convex hull of 
$(d + 1)$ points in a space of at least $d$ dimensions. For example, a first order simplex can be visualized as a zero 
dimensional point, a second order simplex---as a line segment with a vertex at each end, a third order simplex---as a 
triangle with three vertices, etc. (Fig.~\ref{Fig2}A). However, in topological applications that address net properties of 
combinations of simplexes---simplicial complexes---the shapes of the simplexes play no role: the information is contained 
only in the combinatorics of the vertexes shared by the adjacent simplexes. 
This motivates using the so-called ``abstract simplexes''---combinatorial abstractions, defined without any reference 
to geometry, simply as sets of $(d +1)$ elements of arbitrary nature. Thus, abstract simplexes and simplicial complexes
retain only one basic property of their geometric counterparts: just as the triangles of the tetrahedra include their facets, 
an abstract simplex of order $(d+1)$ includes all its subsimplexes of lower orders. As a consequence, a nonempty overlap 
of a pair of simplexes $\sigma$ and $\sigma'$ is a subsimplex of both $\sigma$ and $\sigma'$ (Fig.~\ref{Fig2}A). 

Previous studies \citep{Curto,Chen,Dabaghian2,Arai,Basso,Hoffman,Babichev2} suggest that the topological theory of 
simplicial complexes provides a remarkably efficient semantics for describing many familiar concepts and phenomena 
of hippocampal physiology, as outlined in the following examples. 

\textbf{Example 1. 
A nerve complex $\mathcal{N}$}. The group of overlapping place fields, $\pi_{i_0} \cap \pi_{i_1} \cap ... \cap \pi_{i_d} 
\neq\emptyset$ produced by the place cells $c_{i_0}, c_{i_1}, ... c_{i_d}$ can be represented by an abstract simplex 
$\sigma = [\pi_{i_0}, \pi_{i_1}, ..., \pi_{i_d}]$; the set of all simplexes produced for a place field map $M_{\mathcal{E}}$ 
thus forms a simplicial complex---the nerve of the cover $\mathcal{N}(M_{\mathcal{E}})$ \citep{Curto,Chen,Dabaghian2}.
Every individual place field then corresponds to a vertex, $\sigma_i$, of $\mathcal{N}(M_{\mathcal{E}})$; each nonempty 
overlap of two place fields, $\pi_i \cap \pi_j \neq \emptyset$, contributes a link $\sigma_{ij} \in\mathcal{N}(M_{\mathcal{E}})$, 
a nonempty overlap of three place fields, $\pi_i \cap \pi_i\cap \pi_k \neq\emptyset$, contributes a facet $\sigma_{ijk} \in
\mathcal{N}(M_{\mathcal{E}})$, and so forth. The Alexandrov-\v{C}ech theorem \citep{Alexandroff,Cech} states that if the 
overlapping regions are contractible in $\mathcal{E}$ (i.e., can be continuously retracted into a point), then $\mathcal{N}
(M_{\mathcal{E}})$ and $\mathcal{E}$ have the same number of holes, loops and handles in different dimensions---mathematically, 
they have the same homologies, $H_{\ast}(\mathcal{N}(M_{\mathcal{E}})) = H_{\ast}(\mathcal{E})$. Thus, the nerve 
complex may serve as a schematic representation of the topological information contained in the place field map 
$M_{\mathcal{E}}$ \citep{Babichev1}.

\textbf{Example 2. The coactivity complex $\mathcal{T}$}. 
In the brain, the information is represented via temporal relationships between spike trains, rather than artificial geometric 
constructs such as place fields. However, the place cell spiking patterns can also be described in terms of a simplicial 
``coactivity'' complex $\mathcal{T}(M_{\mathcal{E}})$, which may be viewed as an implementation of the nerve complex 
$\mathcal{N}(M_{\mathcal{E}})$ in the temporal domain. In this construction, every active place cell $c_i$ is represented 
by a vertex, $\sigma_i$, of $\mathcal{T}(M_{\mathcal{E}})$; each coactive pair of cells, $c_i$ and $c_j$, contributes a link 
$\sigma_{ij} = [c_i, c_j]\in\mathcal{T}(M_{\mathcal{E}})$, a triplet of coactive cells contributes a facet 
$\sigma_{ijk} = [c_i, c_j, c_k]\in\mathcal{T}(M_{\mathcal{E}})$, and so forth. 
As a whole, the coactivity complex $\mathcal{T}$ represents the entire pool of the coactive place cell combinations. 
Numerical simulations carried out in \citep{Dabaghian2,Arai,Basso,Hoffman} demonstrate that if the parameters of place 
cells' spiking fall into the biological range, then $\mathcal{T}(M_{\mathcal{E}})$ faithfully represents the topology of two- 
and three-dimensional environments and serves as a schematic representation of the information provided by place cell 
coactivity (Fig.~\ref{Fig2}B). 

\textbf{Example 3. 
Cell assembly complex $\mathcal{T}_{CA}$}. Physiologically, not all combinations of coactive place cells are detected and 
processed by the downstream networks. Therefore, in order to describe only the physiologically relevant coactivities, one 
can construct a smaller ``cell assembly complex'' $\mathcal{T}_{CA}(M_{\mathcal{E}})$, whose maximal simplexes 
represent the combinations of cells that comprise the actual cell assemblies (Fig.~\ref{Fig2}C). Such a complex plays two 
complementary roles: first, it schematically represents the architecture of the cell assembly network (i.e., defines explicitly 
which cells group into which assemblies) and second, it represents the information encoded by this network and hence 
serves as a schematic model of the cognitive map \citep{Babichev2}. 

Previous studies \citep{Dabaghian2,Arai,Basso,Hoffman} concentrated on the lower dimensions ($D \leq 3$) of the 
coactivity and of cell assembly complexes used to represent spatial information, whereas the higher dimensions were 
not addressed or physiologically interpreted. However, a schematic representation of both spatial and nonspatial 
memories should include the full scope of relationships encoded by the cell assemblies; we will therefore use the full 
coactivity complex $\mathcal{T}_{CA}(M_{\mathcal{E}})$ to model a multidimensional memory space.

\textbf{A constructive approach to topology and continuity}. We now make a short mathematical digression to outline the 
key notions necessary for discussing the topology of memory spaces. In general, defining a topological space requires two 
constituents: a set $X$ of spatial primitives---the ``building blocks of space,'' and a set of relationships between them, which 
define spatial order and spatial connectivity. In the standard approach, the topological spaces are comprised of an infinite 
amount of infinitesimal points, and a framework of proximity and remoteness relationships emerges as a matter of combining 
these points into ``topological neighborhoods'' (see Section~\ref{section:methods}). Such system of neighborhoods is referred 
to as a topology on $X$, which we will denote as $\Omega(X)$. In order for the neighborhoods to be mutually consistent, it is 
required that their unions and finite intersections should also be neighborhoods from $\Omega(X)$ (so-called Hausdorff axioms, 
see Section~\ref{section:methods}). Once a consistent framework of neighborhoods is defined, the elements of the set $X$ can 
be viewed as ``spatial locations'' and the set $X$ itself---as a topological space. For example, the environment $\mathcal{E}$, 
viewed as a domain of Euclidean space, contains a continuum of infinitesimal points with Cartesian coordinates $(x,y)$. The 
standard selection of topological neighborhoods in this case is the set of open balls of rational radii, centered at the rational 
points, and their combinations, which define the Euclidean topology $\Omega_{E}(\mathcal{E})$, used in calculus and in 
standard geometries \citep{Alexandrov}.

Modeling a ``memory space'' requires modifying this approach in two major aspects. First, since a memory space emerges 
from the spiking activity of a finite number of neurons, it must be modeled as \emph{finite topological space} 
\citep{Alexandroff,Stong,McCord}, i.e., as a space that may contain only a finite number of elementary locations. Second, since 
every location is encoded by a finite ensemble of place cells, each one of which represents an extended region, the ``spatial 
primitives'' in memory space must be finite domains, rather than infinitesimal points. The latter approach underlies the so-called 
pointfree (or ``pointless'') topologies, geometries \citep{Weil,Johnstone,Laguna, Sambin1,Roeper} and mereotopologies 
\citep{Cohn1,Cohn2}, in which finite regions are considered as the primary objects, whereas the points appear as secondary 
abstractions. As discussed below, these approaches provide suitable frameworks for modeling the biological mechanisms of 
spatial information processing. 

\textbf{A simplicial schema of a memory space}. To build a model of a memory space, we start by noticing that simplicial 
complexes themselves may be viewed as topological spaces, because the relationships between simplexes in a simplicial 
complex $\Sigma$ naturally define a set of topological proximity neighborhoods. Indeed, a neighborhood of a simplex $\sigma$ 
is formed by a collection of simplexes that include $\sigma$ (Fig.~\ref{Fig2}A). It can be verified that the unions and the 
intersections of so-defined neighborhoods satisfy the Hausdorff axioms and hence that any simplicial complex $\Sigma$ 
may be viewed as a finitary topological space $\mathcal{A}(\Sigma$) (see Section~\ref{section:methods}). In mathematical 
literature, such spaces are referred to as Alexandrov spaces, after their discoverer, P. S. Alexandrov \citep{Alexandroff}, 
which motivates our notation.

Importantly, the construction of Alexandrov spaces applies to ``abstract'' simplicial complexes, whose simplexes may 
represent collections of elements of arbitrary nature and hence possess a great contextual flexibility. In our model, individual 
simplexes represent combinations of coactive place cells, believed to encode memory episodes. We may therefore view the 
pool of coactive neuronal combinations as a topological space from two perspectives. On the one hand, one can consider a 
formal ``space of coactivities'' $\mathcal{A}_{\mathcal{E}}({\mathcal{T}_{CA}})$ defined, as the corresponding coactivity 
complexes, in terms of the neuronal spiking parameters. On the other hand, assuming that the combinatorial relationships 
between groups of coactive cells capture relationships between the corresponding memory episodes, one may view the 
collection of memories \emph{represented} by these neuronal activity patterns as elements of a topological \emph{memory 
space} $\mathcal{M}_{\mathcal{E}}({\mathcal{T}_{CA}})$. 
In other words, one can view the Alexandrov space $\mathcal{A}_{\mathcal{E}}({\mathcal{T}_{CA}})$ as a model of the 
memory space $\mathcal{M}_{\mathcal{E}}({\mathcal{T}_{CA}})$ induced by the corresponding cell assembly network. In 
particular, such model can be used to connect the physiological parameters of the latter and the topological characteristics of 
$\mathcal{M}_{\mathcal{E}}({\mathcal{T}_{CA}})$, as we discuss below. Since all subsequent analyses are carried out only
for the memory spaces induced from cell assembly complexes, we will suppress the reference to $\mathcal{T}_{CA}$ in the 
memory space notation.

We would like to note here, that since the simplexes are not structureless objects (i.e., one combination of coactive cells 
represented by simplex $\sigma_1$ may overlap with another combination, represented by a simplex $\sigma_2$, yielding 
a third combination/simplex $\sigma_3$), they represent extended regions, rather than structureless points. 
As a result, the memory space $\mathcal{M}_{\mathcal{E}}$ naturally emerges as a region-based, or ``pointfree'' space, 
in which individual memory episodes correspond to finite regions. Nevertheless, one can easily construct a conventional, i.e., 
point-based, topological space in which a finite set of elementary locations---the ``points''---is organized into the same 
system of proximity neighborhoods as its region-based counterpart (see Section~\ref{section:methods}). In this construction, 
the ``elementary locations'' are simply the smallest regions of $\mathcal{M}_{\mathcal{E}}$, i.e., the ones that cannot be 
further subdivided using the information contained in the place cell coactivity---the ``nodes of the memory space,'' in terminology 
of \citep{Eichenbaum1}. 
In the spatial context, they correspond to the atomic, indecomposable regions. For example, in a mini-memory space encoded 
by two place cells may contain a few ``atomic'' regions: e.g., the region marked by the activity of first, but not the second cell, 
the region marked by the coactivity of both cells and the region marked by the activity of the second, but not the first cell 
(Fig.~\ref{Fig1}A and Figure 12.1 in \citep{Munkres}). In the following we will discuss the organization of such regions in order 
to establish important properties of the memory spaces, e.g. a continuous mapping of the environment $\mathcal{E}$ into a 
memory space $\mathcal{M}_{\mathcal{E}}$.

\section{Results}
\label{section:results}

\textbf{Continuity in memory space}. The discrete memories that comprise a memory space may be triggered by constellations 
of cues and/or actions, that drive the activity of a particular population of cell assemblies \citep{Buzsaki2}. Activation of one cell 
assembly may excite adjacent cell assemblies that represent overlapping memory elements. Thus, as the animal navigates the 
environment, the cell assemblies ignited along a path $\gamma$ form an ``activity packet'' that moves across the network
\citep{Samsonovich,Touretzky,Romani}. If the cell assembly network is represented by a complex $\mathcal{T}_{CA}$, this packet 
is represented by a group of ``active'' simplexes that moves across $\mathcal{T}_{CA}$, tracing a simplicial path $\Gamma$ 
(Fig.~\ref{Fig2}B). As discussed in \citep{Dabaghian2,Arai,Basso,Hoffman,Dabaghian1}, the structure of the simplicial paths captures 
the shape of the corresponding physical paths and hence represents the connectivity of the environment. 
For example, a contractible simplicial path corresponds to a contractible physical rout, whereas a non-contractible simplicial path 
marks a non-traversable domain occupied by an obstacle, e.g., by a physical obstruction or by a predator (Fig.~\ref{Fig2}B,C).

Intuitively, one would expect that a continuous physical trajectory should be represented by a ``continuous succession'' of activity 
regimes of the place cells that represents a continuous succession of memory episodes. Indeed, the topological structure of the 
memory space provides a concrete meaning for this intuition. It can be shown that the environment $\mathcal{E}$ maps continuously 
into the memory space $\mathcal{M}_{\mathcal{E}}$, and in particular, that each continuous trajectory $\gamma$ traced by the 
animal in the physical environment maps into a continuous path $\wp$ in the memory space $\mathcal{M}_{\mathcal{E}}$ (see
Section~\ref{section:methods}). It should be noted however, that these are different continuities: the physical trajectory $\gamma$ 
is continuous in the Euclidean topology of the environment, whereas the path $\wp$ is continuous in the topology of the memory space. 
This distinction is due to fact that the environment $\mathcal{E}$ and the memory spaces $\mathcal{M}_{\mathcal{E}}$ are not 
topologically equivalent to each other: one can map the rich Euclidean topology onto the discrete finite topology of a memory space, but 
not vice versa. In other words, despite the continuity of mapping from $\mathcal{E}$ into $\mathcal{M}_{\mathcal{E}}$, the memory 
space remains only a discretization of the environment, which nevertheless serves as a topological representation of $\mathcal{E}$ and 
can be continuously navigated.

\textbf{Topological properties of memory spaces} can be studied from two perspectives: from the perspective of algebraic topology 
that captures the large-scale structure of $\mathcal{M}_{\mathcal{E}}$ in terms of topological invariants \citep{Munkres}, or from
the perspective of the so-called general topology \citep{Alexandrov}, which describes the topological ``fabric'' of $\mathcal{M}_{\mathcal{E}}$, 
in terms of the proximity neighborhoods.

The algebraic-topological properties of the coactivity complexes were studied in \citep{Dabaghian2,Babichev2,Babichev3,Babichev4}. 
There it was demonstrated that if place cell populations operate within biological parameters, then the number of topological loops 
in different dimensions of the coactivity complex---the Betti numbers $b_n(\mathcal{T}_{CA})$ \citep{Munkres}---match the Betti 
numbers of the environment $b_n(\mathcal{E})$. Moreover, the correct shape of the coactivity complex emerges within a biologically 
plausible period that was referred to as learning time, $T_{\min}$. These results apply directly to the memory spaces, since the Betti 
numbers of a memory space $\mathcal{M}_{\mathcal{E}}$ are identical to those of the coactivity complex $\mathcal{T}_{CA}$ that 
produced it \citep{Alexandroff}. (For a mathematically oriented reader, we mention that the homological structure of 
$\mathcal{M}_{\mathcal{E}}$ should be defined in terms of singular homologies, whereas the structure of the coactivity complex is 
described in terms of simplicial homologies. However, for the cases considered below, these homologies coincide, so we omit the 
discussion of the differences \citep{McCord}). This implies, in particular, that the memory space that correctly represents the topology 
of the environment emerges together with the corresponding coactivity complex during the same learning time $T_{\min}$, for the 
same set of spiking parameters (in terminology of \citep{Dabaghian2}, within the ``learning region,'' $\mathcal{L}$). 

\begin{figure} 
\includegraphics[scale=0.89]{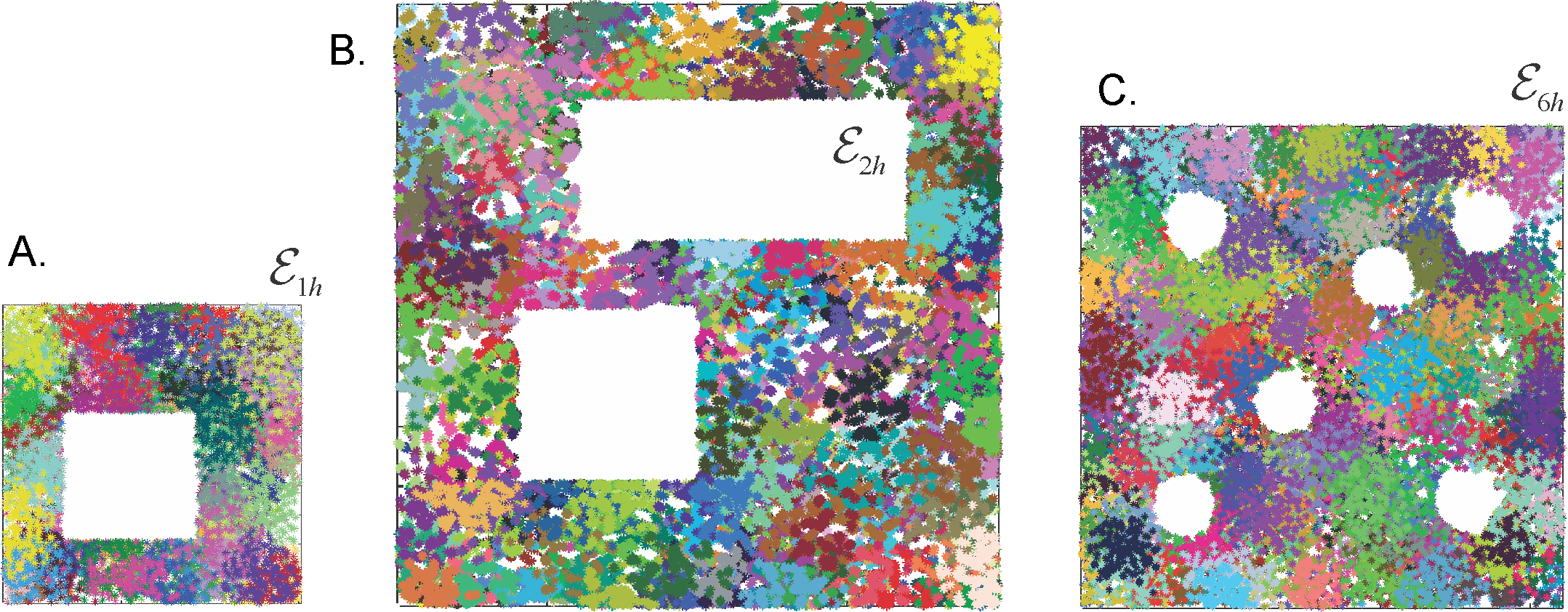}
\caption{{\footnotesize \textbf{Place field maps in three simulated environments}. ({\bf A}.) An example of a place field map 
simulated in a $1m \times 1m$ environment with one hole in the middle,  $\mathcal{E}_{1h}$, that was previously studied in 
\citep{Dabaghian2,Basso,Babichev2}. Dots of different colors represent spikes produced by different place cells. Clusters of 
dots represent the corresponding place fields. ({\bf B}). A place field map simulated in $2m \times 2m$ environment with two 
holes, $\mathcal{E}_{2h}$, studied in \citep{Arai}. ({\bf C}.) The third environment $\mathcal{E}_{6h}$ ($1.6m \times 1.6m$
in size) is similar to the behavioral arena studied in \citep{Tse1}, where the concept of the Morris' schemas was introduced. 
Ten different place field maps were simulated in each environment and used to produce a cell assembly network, as described 
in \citep{Babichev2}. The mean size of the place fields (20 cm) and the mean firing rate of the place cells (14 Hz) is the same 
in all cases.}} 
\label{Fig3}
\end{figure} 

Importantly, the learning times and other global characteristics of $\mathcal{T}_{CA}$ produced via algebraic topology 
techniques are insensitive to many details of the place cell spiking activity \citep{Dabaghian2,Babichev2,Babichev3,Babichev4}. 
For example, the learning time $T_{\min}$ depends mostly on the mean place field sizes and the mean peak firing rates, but 
it does not depend strongly on the spatial layout of the place fields or on the limited spiking variations. 
The question arises, how sensitive is the ``fabric'' of the memory space to the parameters of neuronal activity? 

To address this question we simulated ten different place field maps $M_i$, $i=1,...,10$, in three environments (Fig.~\ref{Fig3}), 
and verified that the corresponding nerves $\mathcal{N}_{\mathcal{E}}(M_i)$, coactivity complexes $\mathcal{T}(M_i)$ and cell 
assembly complexes $\mathcal{T}_{CA}(M_i)$ produced the required large-scale topological characteristics (i.e., the same Betti 
numbers: $b_0(\mathcal{E}_{1h})=b_0(\mathcal{E}_{2h})=_0(\mathcal{E}_{6h})=1$, $b_1(\mathcal{E}_{1h})=1$, 
$b_1(\mathcal{E}_{2h})=2$, $b_0(\mathcal{E}_{6h})=6$, and $b_n(\mathcal{E}_{1h})=b_n(\mathcal{E}_{2h})=
b_n(\mathcal{E}_{6h})=0$, $n\geq 2$).
We then built and analyzed the memory spaces for the cell assembly complexes, and analyzed their general-topological structure. 
Mathematically, the discrete topology of an Alexandrov space can be represented by a numerical matrix---the Stong matrix 
$S_{\mathcal{A}}$, which enables effective numerical analyses (see Section~\ref{section:methods} and \citep{Stong}). 
Analyzing the Stong matrices for $\mathcal{M}_{1h}$, $\mathcal{M}_{2h}$ and $\mathcal{M}_{6h}$, we observed the memory 
spaces constructed for different place field maps in the same environment have different topologies. In other words, a memory space 
$\mathcal{M}_{\mathcal{E}}(M_i)$ encoded by a cell assembly network that corresponds to the place field map $M_i$ cannot, 
in general, be continuously deformed into the memory space $\mathcal{M}_{\mathcal{E}}(M_j)$, that corresponds to place 
field map $M_j$ in the same environment. 
From the mathematical perspective, this outcome is not surprising: since memory spaces are topologically inequivalent to the 
environment (a continuous mapping $\mathcal{E}\to \mathcal{M}$ exists but the continuous mapping $\mathcal{M} \to\mathcal{E}$ 
does not), two different memory spaces produced in the same environment may be inequivalent to each other. However, from a 
neurophysiological perspective, these results imply that a memory space reflects not only the large-scale topological structure of the 
environment, but also the specifics of a particular place field map, e.g., local spatial relationships between individual place fields. 

\begin{figure} 
{\includegraphics[scale=0.84]{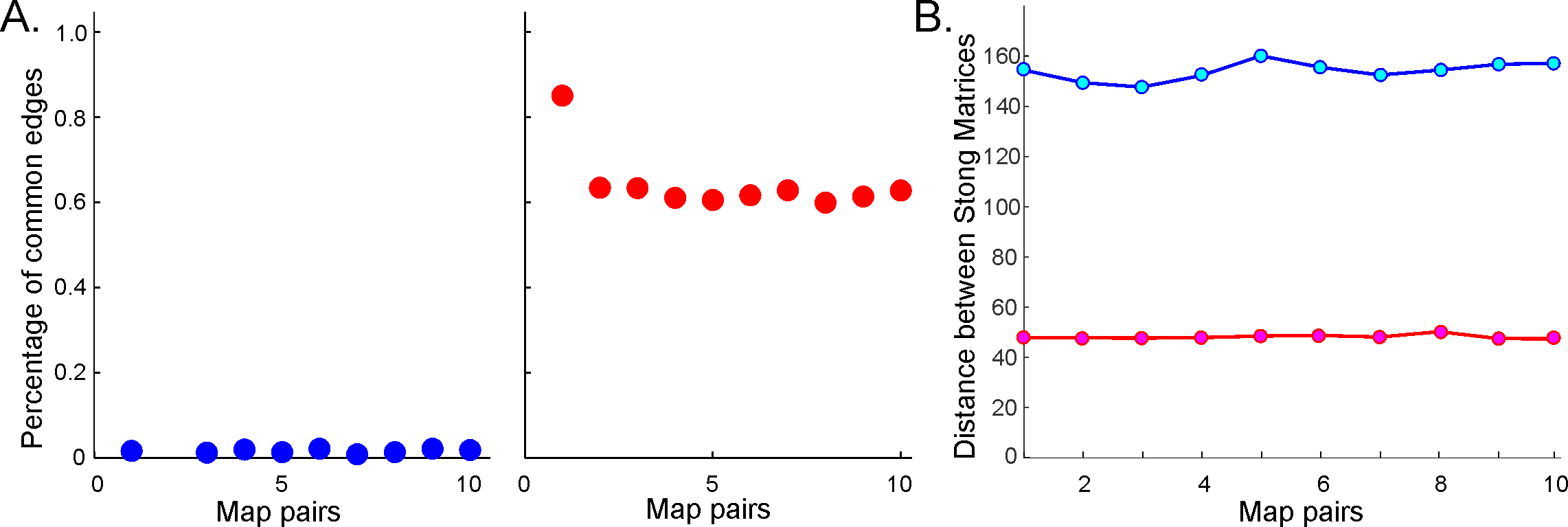}}
{\caption{{\footnotesize \textbf{Similarity between memory spaces and place field remapping}. ({\bf A}) Proportion of one-dimensional
simplexes (links) shared by ten pairs of coactivity complexes, $\mathcal{T}_{CA}(M_i)$ and $\mathcal{T}_{CA}(M_j)$, induced 
from ten pairs of place field maps in the six-hole environment $\mathcal{E}_{6h}$. Left panel illustrates the case in which the 
centers of the place fields in $M_i$ and $M_j$ are independently scattered (global remapping); right panel illustrates the case 
in which the place field positions are fixed, but the place cells' firing rates and place field sizes are altered by 5\% (rate remapping). 
In the latter case, most links are preserved, implying that the one-dimensional ``skeleton'' of the coactivity complex \citep{Munkres} 
(or the corresponding coactivity graph $\mathcal{G}$ \cite{Babichev1}) is largely preserved in rate remappings. ({\bf B}) The distance 
norms between the Stong matrices in both global (blue) and rate (red) remappings are significant, implying that the corresponding 
memory spaces $\mathcal{M}_{6h}(M_i)$ and $\mathcal{M}_{6h}(M_j)$ are topologically distinct (see Methods). However, the 
change of the memory space's topology in rate remapping is smaller than in global remapping.}}
\label{Fig4}}
\end{figure}

Further analyses point out that even if the place field map is geometrically the same but the firing rates change by less than 5\%, 
the cell assembly networks built according to the methods outlined in \citep{Babichev2} also change. As a result, the corresponding 
memory spaces come out to be topologically distinct from one another, although the differences between their respective Stong 
matrices are smaller than the differences between the Stong matrices induced by the different maps place field maps (Fig.~\ref{Fig4}).
 
These results can be physiologically interpreted in the context of the so-called place field remapping phenomena, which we briefly 
outline as follows. As mentioned in the Introduction, if the changes in the environment are gradual, then the relative order of the 
place fields in space remains the same and place cells exhibit only small changes in the frequency of spiking \citep{Colgin2,Dupret}. 
In contrast, if an environment is changed abruptly, e.g., if major cues suddenly appear or disappear, then the place cells may 
independently shift the locations of their place fields across the entire environment and significantly change their firing rates, i.e., 
one place field map is substituted by another \citep{Geva,Kammerer2,Fyhn}. 
The former phenomenon, known as \emph{rate} remapping, is believed to represent variations of contextual experiences embedded 
into a stable spatial code, while latter, the \emph{global} remapping, is believed to indicate a restructuring of cognitive representation 
of the environment. This is confirmed by our model: the differences between the memory spaces produced by two geometrically 
distinct place field maps $M_i$ and $M_j$ (physiologically, one can view a place field map $M_j$ as a result of a remapping from a 
map $M_i$) are large, whereas rate remapping produces much smaller variations in the structure of the memory space (Fig.~\ref{Fig4}). 
In either case, the corresponding memory spaces are continuous images of the environment (i.e., a continuous mapping $\mathcal{E}
\to\mathcal{M}_{\mathcal{E}}$ exists in all cases) and $\mathcal{M}_{\mathcal{E}}$ can be continuously navigated, see Suppl. Movies
\citep{SupMov1,SupMov2,SupMov3}. In particular $\mathcal{M}_{\mathcal{E}}$ always correctly represents the large-scale topology 
of the environment (the Betti numbers $b_n(\mathcal{E})$ and $b_n(\mathcal{M}_{\mathcal{E}})$ match for all $n$s).

\textbf{Reduction of the memory spaces}. Over time, the memory frameworks undergo complex changes: detailed spatial 
memories initially acquired by the hippocampus become coarser-grained as they consolidate into long-term memories stored 
in the cortex \citep{Rosenbaum,Hirshhorn,Preston,Winocur}. From the memory space's properties perspective, this suggests 
that a memory space associated with a particular memory framework (e.g., with a particular environment) looses granularity
but preserves its overall topological structure. The physiological mechanisms underlying these processes and the theoretical 
principles of memory consolidation are currently poorly understood and remain a matter of debate \citep{OReilly,Benna}. 
However, the topological framework proposed above allows an impartial, schematic description of consolidating the topological 
details in memory spaces and producing a more compact representations of the original memory framework.

\begin{figure} 
\includegraphics[scale=0.89]{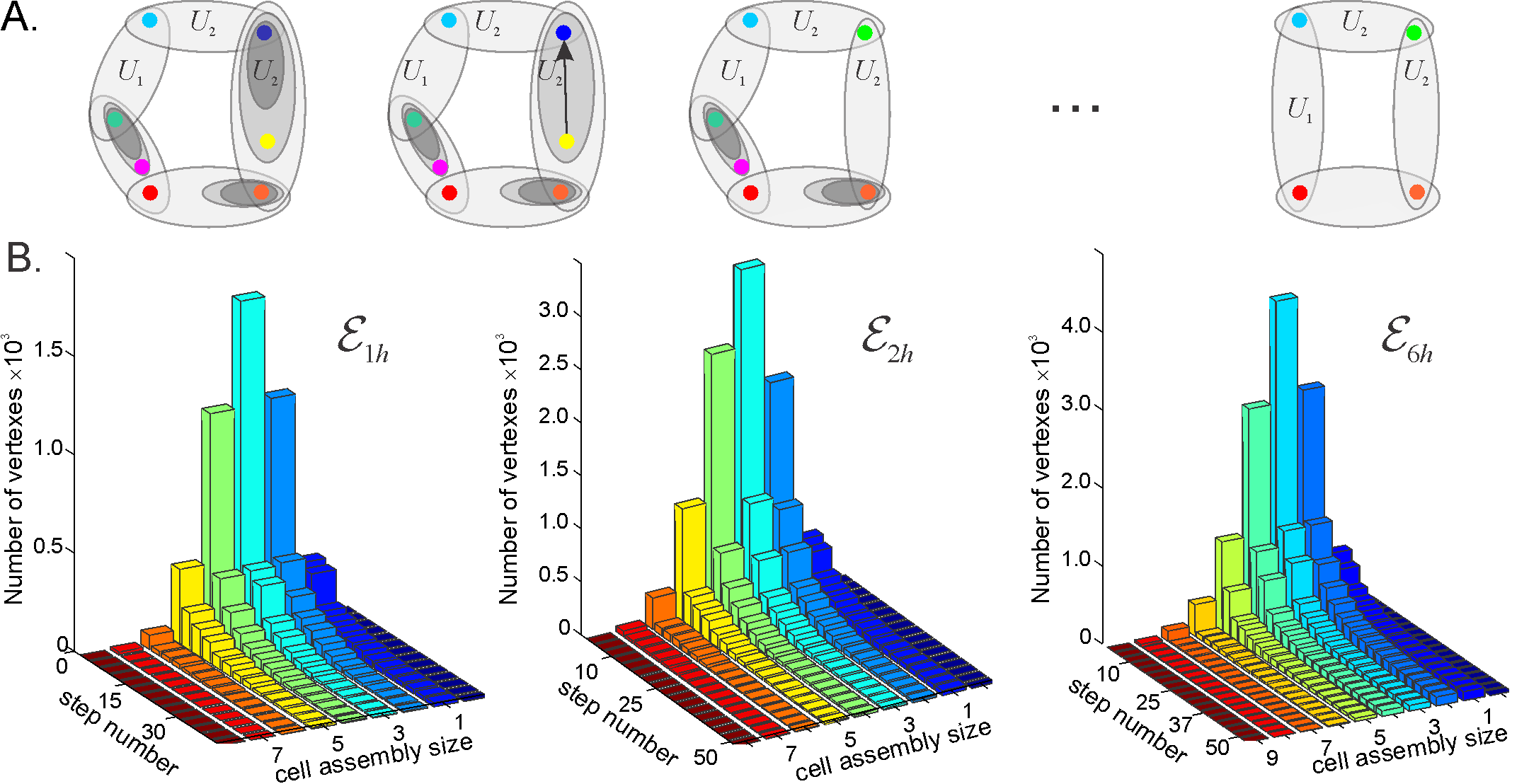}
\caption{{\footnotesize \textbf{Reduction of finite topological spaces}. ({\bf A}) A finite topological space containing seven points, 
with the topological neighborhoods shown by the ovals. Note, that the blue and the yellow point on the first panel are separated 
by a single neighborhood. If this neighborhood is removed (middle panel), then the yellow point collapses onto the blue point. 
The resulting green point represents a single ``combined'' location. The minimal possible topological construction with circular 
topology contains four points.
({\bf B}) The reduction of the number of points in three memory spaces, $\mathcal{M}_{1h}(M)$, $\mathcal{M}_{2h}(M)$ 
and $\mathcal{M}_{6h}(M)$, in the three environments shown on Fig.~\ref{Fig3}, as a function of the reduction step. As the 
topology is consolidated, the number of simplexes--and of the corresponding points--drops from thousands to a few dozens 
(see Fig.~\ref{Fig4}). Note that the dimensionality of the original simplexes ranges between $D=7$ for $\mathcal{T}_{1h}$ 
and $D=9$ for $\mathcal{T}_{6h}$, whereas most elements in the reduced spaces have dimensionality $D\approx 3$. Thus, 
the higher order memory combinations are consolidated into smaller-dimensional framework.}}
\label{Fig5}
\end{figure}

As mentioned in the Section~\ref{section:model}, topological neighborhoods define proximity and remoteness between spatial 
locations. However, certain neighborhoods may carry only limited topological information. For example, if a neighborhood $U_i$ in 
a space $\mathcal{A}$ is entirely contained in a single larger neighborhood $U_k$ and is involved in the same relationships with 
other neighborhoods as $U_k$, then it only adds granularity to the topology of $\mathcal{A}$ without affecting its overall structure 
(Fig.~\ref{Fig5}A). In such case, the topology $\Omega(\mathcal{A})$ can be coarsened by removing $U_i$ and producing a 
``reduced'' space $\mathcal{A}'$ that is topologically similar to $\mathcal{A}$ (homotopically equivalent, see 
Section~\ref{section:methods} and \citep{Stong,Osaki,McCord}). If such coarsening procedure is applied multiple times, then the 
resulting chain of transformations, $\mathcal{A}\to\mathcal{A}' \to\mathcal{A}''\to...\to\mathcal{A}^{(n)}$, generates a sequence 
of progressively coarser and coarser spaces that retain the homological identity of $\mathcal{A}$ (e.g., same Betti numbers). 

\begin{figure} 
\includegraphics[scale=0.89]{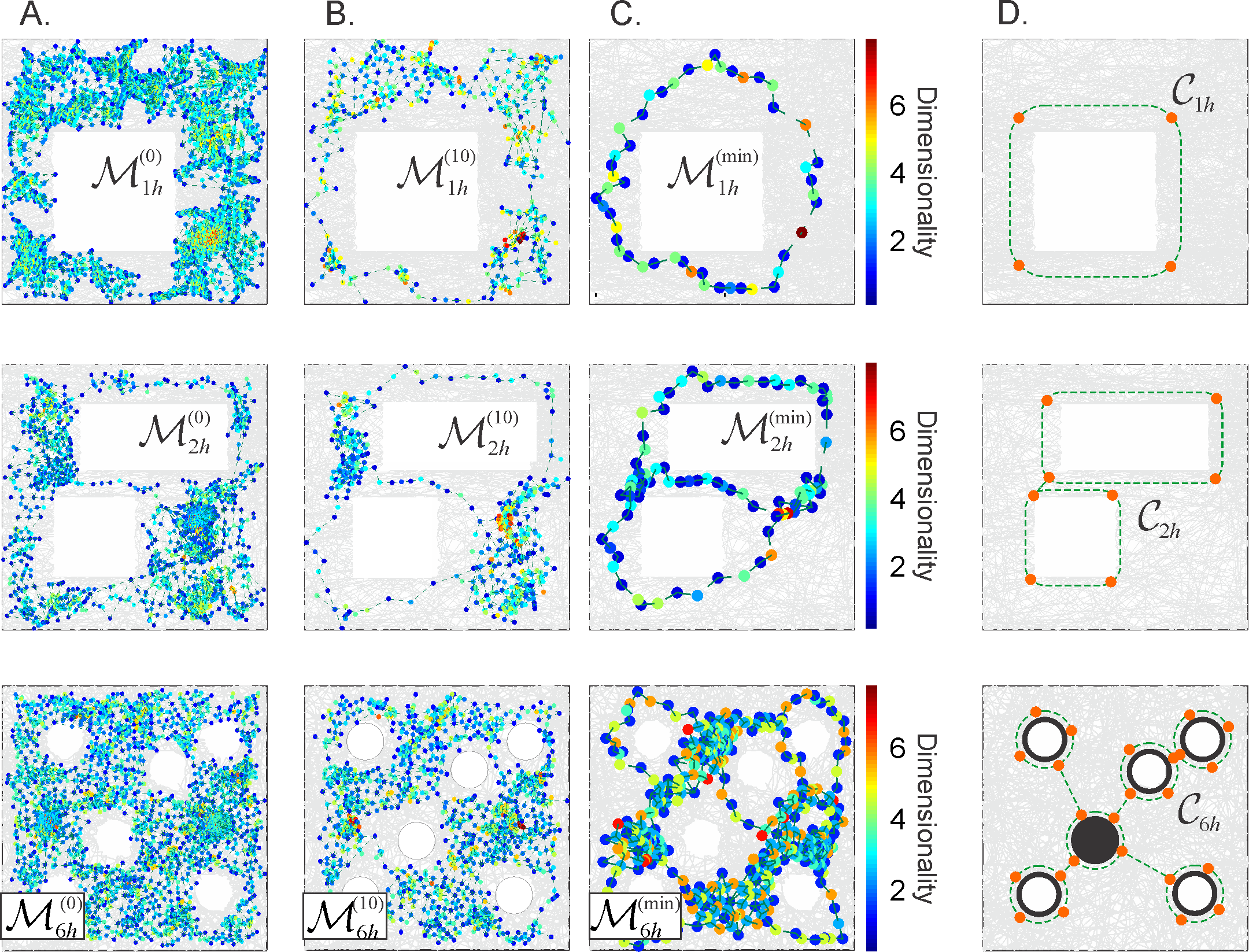}
\caption{{\footnotesize \textbf{Reduction of the Alexandrov spaces into their cores and the corresponding Morris' schemas.}. 
({\bf A}) Points of the memory space, induced from a cell assembly coactivity complex, constructed for the place field maps
shown on Fig.~\ref{Fig3}. The color of points corresponds to the dimensionality of the corresponding simplexes. 
({\bf B}). The points of the memory spaces after 10 reduction steps (left column) and ({\bf C}) the points of the topological 
cores of the memory spaces, obtained after a maximal reduction of the memory spaces (right column). For more examples 
see Suppl. Fig.~1. 
({\bf D}). The minimal cores that correspond to each environment. A four point core on the top panel provides a minimal 
topological representation of a circle, the two linked four point circles represent the environment with two holes (middle panel). 
In the case of the environment considered in the experiment discussed in \citep{Tse1}, Fig.~\ref{Fig3}D, the minimal core 
corresponds to the Morris' schema.}}
\label{Fig6}
\end{figure}

To the extent to which the consolidated memory frameworks retain the structure of the memory space $\mathcal{M}_{\mathcal{E}}$, 
they can be interpreted as its topological reductions. Thus, in the proposed approach, the consolidation process may be modeled via a 
sequence of less granular and more compact memory spaces, $\mathcal{M}_{\mathcal{E}}\to \mathcal{M}'_{\mathcal{E}} \to
\mathcal{M}''_{\mathcal{E}} \to...\to\mathcal{M}_{\mathcal{E}}^{(n)}$ as discussed in \citep{Stong,Osaki,McCord}, see 
Fig.~\ref{Fig6}A-C and Suppl. Movies~\citep{SupMov4}). 

Importantly, the reduced memory spaces $\mathcal{M}_{\mathcal{E}}^{(k)}$ remain continuous images of both the original 
memory space $\mathcal{M}_{\mathcal{E}}$ and of the environment $\mathcal{E}$. However, unlike the full memory space, 
the reduced memory spaces are not just ``topological replicas'' of the cell assembly complex: as the memory space is reduced, the 
direct correspondence between the simplexes of $\mathcal{T}_{CA}$ and the elements of $\mathcal{M}^{(k>0)}_{\mathcal{E}}$ 
disappears. The reduction of neighborhoods and points in $\mathcal{M}^{(k>0)}_{\mathcal{E}}$ corresponds to eliminating certain 
simplexes of the cell assembly complex $\mathcal{T}_{CA}$, i.e., to a restriction of the processed place cell coactivity inputs. The 
connections required to process these inputs can form a smaller cell assembly network that encodes the consolidated memory space
$\mathcal{M}_{\mathcal{E}}^{(k)}$. 

The smallest memory space obtained at the last step of the reduction process $\mathcal{M}_{\mathcal{E}}^{(\max)}$ (i.e., the 
one that cannot be reduced any further), retains the overall topological properties of the original memory space in the most compact 
form, i.e., using the smallest number of points and neighborhoods obtainable via a particular consolidation process (Fig.~\ref{Fig6}C). 
The exact structure of such an ``irreducible'' memory space, referred to as \emph{core} $\mathcal{C}(\mathcal{M}_{\mathcal{E}}$)
of the memory space $\mathcal{M}_{\mathcal{E}}$, depends on the reduction sequence (\citep{Stong,Osaki,McCord} and Suppl. Fig. 1). 
However, for every environment $\mathcal{E}$, considered as topological space, there exists a unique core $\mathcal{C}_{\mathcal{E}}$ 
(see Fig.~\ref{Fig6}D and \citep{Stong,Osaki}), which schematically represents its basic, skeletal structure, approximated by $\mathcal{C}
(\mathcal{M}_{\mathcal{E}}$).

Similar compact, schematic representations of the memory structures are frequently discussed in neurophysiological 
literature. For example, in \citep{Tse1} it was proposed that, as a result of learning, animals may acquire a cognitive 
schema---a consolidated representation of the spatial structure of the environmental and of the behavioral task \citep{Tse2,Morris}. 
Specifically, in the case of the environment $\mathcal{E}_{6h}$ shown 
on Fig.~\ref{Fig3}C, the Morris' schema has the form shown on the bottom panel of Fig.~\ref{Fig6}D, i.e., it is structurally 
identical to the core of $\mathcal{E}_{6h}$. 
We use this observation to suggest that the Morris' schemas may in general be identified with the cores of the memory 
spaces produced by a particular cell assembly network in a given environment, and that acquiring a Morris' schema through 
a memory consolidation process may be modeled as the memory space reduction. 

Under such hypothesis, the model allows computing specific Morris' schemas from their respective memory spaces, using 
the physiological parameters of neuronal activity and the corresponding cell assembly network architecture. Specifically, one 
can identify the number of elements in a given schema, their projected locations in the environment and their shapes. 
For the memory spaces constructed for different place field maps of the environments shown on Fig.~\ref{Fig3}, the computed 
Morris schemas form a set of connected loops encircling the topological obstacles, as 
suggested in \citep{Tse2,Morris}. The density of the nodes along the constructed Morris' schemas (Fig.~\ref{Fig6}C) is 
higher than in heuristic constructions, and similar to the characteristic distance between the place field centers in the
corresponding maps.

\section{Discussion}
\label{section:discussion}

According to the cognitive map concept, spatial cognition is based on internalized representation of space encoded by the 
hippocampal network \citep{Tolman}, which was broadly studied both experimentally and theoretically, in particular, using 
the topological approach \citep{Curto,Dabaghian2,Babichev2,Chen}. Here we extend the topological schema approach proposed 
in \citep{Babichev1}, to describe not only spatial, but also nonspatial memories in a single mathematical construct---a topological 
space with specific mathematical properties, induced by the physiological parameters of neuronal activity. 
The resulting model allows demonstrating, first, that the memory spaces incorporate representations of spatial experiences, 
i.e., that the cognitive maps are naturally embedded into memory spaces. In particular, the latter captures the topological 
structure of the navigated environment, so that the physical trajectories are represented by continuous paths in the memory 
space. Second, the model allows interpreting the hippocampal remapping phenomena in the context of the net topological 
properties of the memory spaces, both from the algebraic and from the general topological perspectives. Lastly, it connects 
the memory space structure to the Morris' schemas, by providing a schematic representation for the memory consolidation 
process.

\textbf{Memory spaces in other topological schemas}. Simplicial coactivity complexes, e.g., the ones discussed 
in the Examples 2 and 3 of Section~\ref{section:model}, are used to represent spatial information by a population of 
readout neurons responding to nearly simultaneous activity of the presynaptic place cells \citep{Babichev1}. 
However, the construction of the memory space discussed above is by no means limited to the particular syntax 
of processing the spiking outputs of the place cells. The key property of a simplicial complex that turns it into a 
space is the partial ordering of its simplexes, produced by the containment relationship: $\sigma_1$ is ``smaller'' 
than $\sigma_2$, if $\sigma_2$ contains $\sigma_1$ (i.e., $\sigma_1 <\sigma_2$ if $\sigma_2\cap\sigma_1 = \sigma_1$). 
However, all topological schemas discussed in \citep{Babichev1} define partial orders, and without going into mathematical 
details, we point out that all partially ordered sets---\emph{posets}---can be viewed as a topological spaces, regardless of 
the nature of the order relationships \citep{Vickers,Davey}. Thus, each topological schema 
$\mathcal{S}$ defines a specific finitary topological space, $\mathcal{M}_{\mathcal{S}}$, which can be interpreted as 
the memory space encoded by the cell assembly network that $\mathcal{S}$ represents. For example, a mereological 
schema $\mathcal{F}$, based on the cover relation, defines partial order ``covered region $x$ is smaller than the covering 
region,'' ($x < y$ iff $x \blacktriangleleft y$). The $\textsf{RCC5}$ schema $\mathcal{R}_{5}$, based on five 
topological relations (partial overlap $\textsf{PO}$, proper part $\textsf{PP}$, its inverse $\textsf{PPi}$, discrete 
$\textsf{DR}$ and equal $\textsf{EQ}$, see Fig.~\ref{Fig1}A and \citep{Cui,Cohn3}) is also partially ordered. 
In this case, a region $x$ is smaller than $y$ if $x$ is a proper part of $y$, or, if two regions $x$ and $y$ partially 
overlap, $\textsf{PO}(x,y)$, then they share a \emph{smaller} region $z$ that is a proper part of both $x$ and $y$, 
i.e., $\textsf{PP}(z,x)$, $\textsf{PP}(z,y)$ \citep{Renz}. The discrete (\textsf{DR}) or equal (\textsf{EQ}) regions 
are unrelated. The posets $\mathcal{P}_{\mathcal{F}}$ and $\mathcal{P}_{\mathcal{R}}$ corresponding to these 
schemas define their respective finitary topological spaces $\mathcal{M}_\mathcal{F}$ and $\mathcal{M}_{\mathcal{R}}$
that represent the topology environment just as the simplicial schema $\mathcal{M}_\mathcal{T}$ discussed above. 

Given the same physiological parameters (e.g., the same number of place cells) the memory spaces produced by 
different schemas may differ from one another, e.g. some of them may have stronger topologies than others. However, 
all memory spaces may be regarded as finitary topological spaces that can be considered on the same footing, irrespective 
of the specific set of rules according to which the information provided by individual place cells is combined in $\mathcal{S}$. 
Thus, the proposed model of memory spaces allows relating the capacity of different cell assembly networks, which may
potentially implement different computational principles for processing spatial information, to represent information. 

\textbf{Intrinsic representation of space}. Current understanding of hippocampal neurophysiology rests on the assumption 
that place cells' spiking ``tags'' cognitive regions. Such approach allows describing the information contained in the spike 
trains phenomenologically, without addressing the ``hard problem'' of how the brain can intrinsically interpret spiking activity 
as ``spatial'' \citep{Chalmers}. It therefore remains unclear in what sense the spiking activity may actually produce a 
``cognitive region,'' in what sense two such regions may ``overlap'' or ``contain one another,'' and so forth. 
Yet, in neuroscience literature it is recognized that ``\emph{allocentric space is constructed in the brain rather than perceived, 
and the hippocampus is central to this construction}'' \citep{OKeefe,Nadel}. Paraphrasing L. Nadel and H. Eichenbaum 
\citep{Nadel}, it remains unclear how can ``spaceless'' data enter the hippocampal system and spatial cognitive maps 
come out. In this connection, we would like to point out that the topological approach discussed above may shed light on 
this problem, by allowing to interpret spatiality in purely relational terms, as a construct emerging from the relationships 
between the signals, implemented by neuronal networks with specific architecture.

\section*{Mathematical and Computational Methods}
\label{section:methods}
Establishing a topological correspondence between the environment and the memory space requires a few definitions.

\begin{figure}[H]
\includegraphics[scale=0.87]{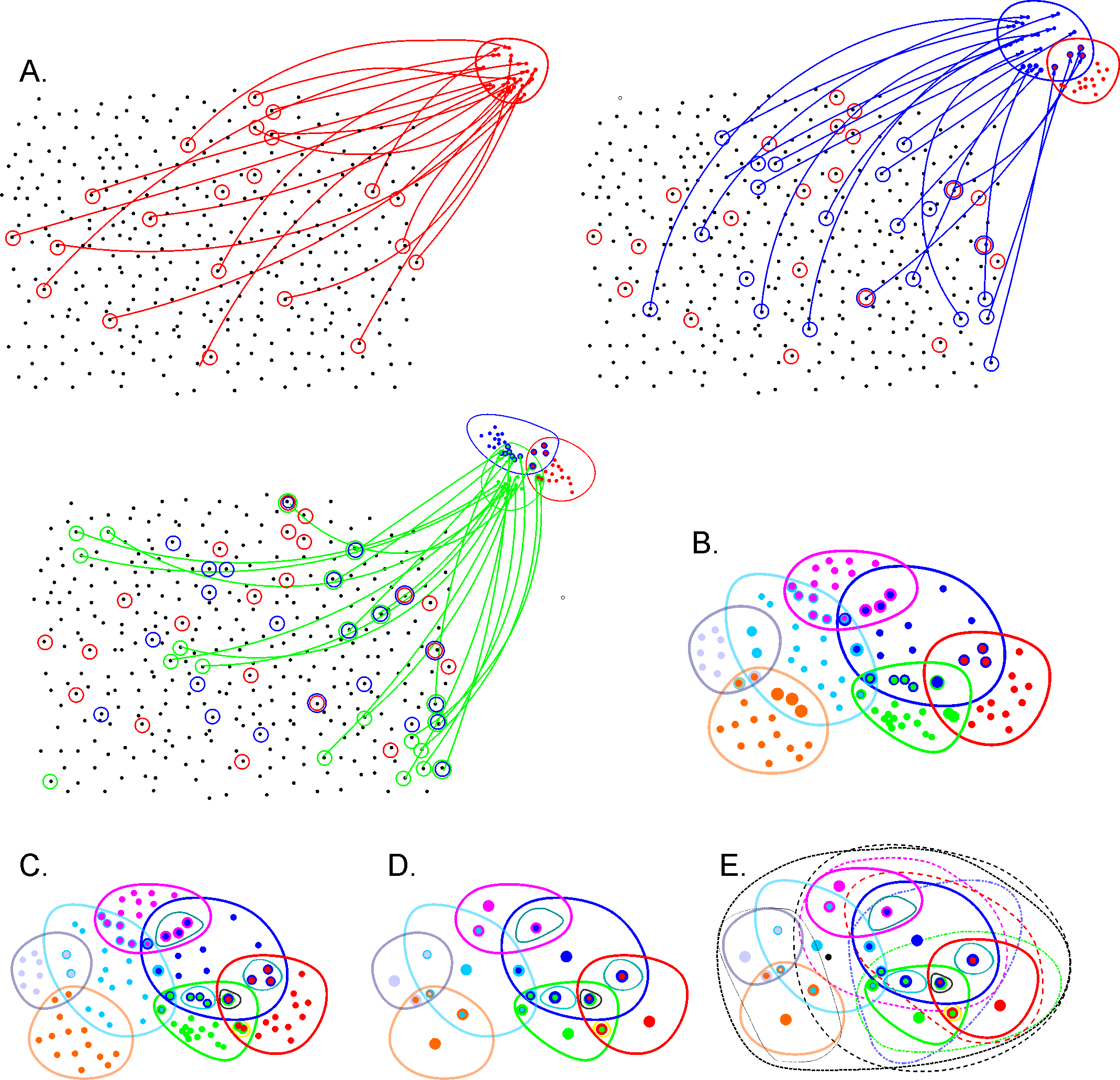}
\caption{{\footnotesize \textbf{Basic notions of point set topology}. ({\bf A}). A set $X$ with no spatial structure turns 
into a topological space as its elements are combined into topological neighborhoods. For example, a subcollection of 
elements of $X$ (marked by red circles) may be selected to form the neighborhood ``red'' points, $U_r$. Another 
collection of elements (blue circles) may form another, ``blue'' neighborhood $U_b$ that may overlap with the red 
neighborhood $U_r$, yet another set may form the green neighborhood $U_g$ and so forth. ({\bf B}). Eventually, the 
elements of $X$ are grouped into a system of neighborhoods---in this case, seven neighborhoods. 
({\bf C}). All intersections between these neighborhoods define a topological base $\mathfrak{B}$---a set of basic 
neighborhoods whose combinations yield arbitrary neighborhoods on $X$. 
({\bf D}). The topology base defines the ``resolution'' of the corresponding topology: if two points share an identical 
system of neighborhoods, then they cannot be separated from each other, or ``resolved'' by the corresponding topology. 
The spaces in which for every two points $x$ and $y$ there is a neighborhood that contains one point but not the other
are referred to as $T_{0}$ spaces. In particular, all Alexandrov spaces are $T_0$-separable. In the illustrated example, 
the topology base can ``resolve'' only 20 points, whereas all other elements of $X$ fuse into these representative 
``locations.'' 
({\bf E}). Adding the unions (only some unions are illustrated by black dashed lines) produces the full system of 
neighborhoods, a finitary topology $\Omega(X)$.}}
\label{Fig7} 
\end{figure}

1. A \emph{topology} on a space $X$ is established by a system $\Omega(X)$ of topological neighborhoods, which obey 
the Hausdorff axioms: any unions and finite overlaps of the topological neighborhoods $U_i \in \Omega(X)$ produce another 
neighborhood from the same system $\Omega(X)$ (Fig.~\ref{Fig7}). The empty set and the full set $X$ also belong to 
$\Omega(X)$ \citep{Alexandrov}.

2. A \emph{topology base} $\mathfrak{B} = {B_i}$ consists of a smaller set of ``base'' neighborhoods that can be 
combined to produce any other neighborhood $U_i$ of $\Omega$. A key property of a topology base is that it is closed 
under the overlap operation: an intersection of any two base neighborhoods yield (or, more generally, contain) another 
base neighborhood. A topology base generates a unique topology for which it forms a base, and hence it is a convenient 
tool for studying topological spaces (a rough analogy is a set of basis vectors in a linear space, \citep{Alexandrov}). 

\emph{Example 1: Euclidean plane}. The standard choice of a topological base $\mathfrak{B}_E$ of a Euclidean domain 
$\mathcal{E}$ are the open balls of rational radii, centered at the points with rational coordinates. Every nonempty 
overlap of a finite collection of such balls contains a ball with a smaller radius. The full set of the topological neighborhoods 
in the resulting topology is given by the arbitrary unions of these balls \citep{Alexandrov}. 

\begin{figure}[H] 
\includegraphics[scale=0.8]{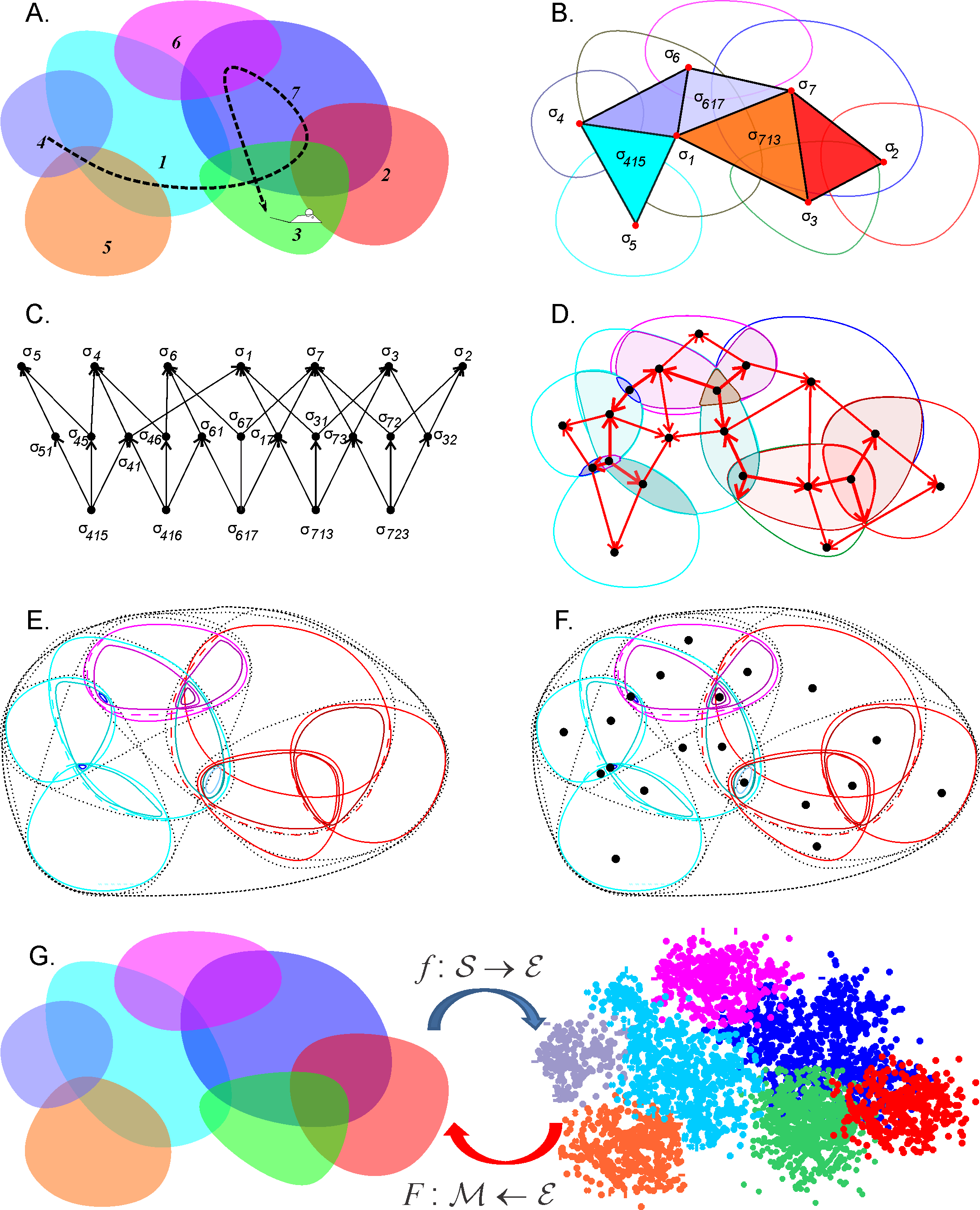}
\caption{{\footnotesize \textbf{Discrete topological spaces and place field maps}. ({\bf A}) A schematic, mini-place 
field map that consists of seven place fields (colored ovaloids), traversed by a fragment of the animal's trajectory 
(dashed line). ({\bf B}) The corresponding nerve complex $\mathcal{N}_7$, which contains topological information 
about the environment. Its vertexes, $\sigma_i$, correspond to the place fields, links $\sigma_{ij}$, to overlapping 
pairs, the triangles $\sigma_{ijk}$ to simultaneously overlapping triples of place fields. Alternatively, one can view 
this as the coactivity complex $\mathcal{T}_7$, whose vertexes correspond to active place fields, links to pairs of
coactive cells, triangles to coactive triples of cells, etc. 
({\bf C}) The partially ordered set--\emph{poset} $\mathcal{P}_7$ corresponding to the nerve $\mathcal{N}_7$. 
({\bf D}) The simplexes of the simplicial complex $\mathcal{N}_7$ (or the elements of the poset $\mathcal{P}_7$) 
map into the atomic elements of the place field map. 
({\bf E}) The poset $\mathcal{P}_7$ can be viewed as a pointfree (relational) space built from the regions defined
by the place cell (co)activity. 
({\bf F}). The corresponding point-based Alexandrov space should be viewed as an analogue of Fig.~\ref{Fig1}B. 
({\bf G}) A spatial mapping from the memory map to the environment and the continuous mapping from the 
environment into memory space, $\mathcal{M}_{\mathcal{T}}$.}}
\label{Fig8}
\end{figure}

\emph{Example 2: Cover induced topologies}. One can generate an alternative topology for the Euclidean domain 
$\mathcal{E}$ by covering it by a set of regions $U_i$ and by augmenting this set with the regions obtained by all 
possible intersections $U_i \cap U_j \cap... \cap U_k$. By construction, the resulting system of regions will be closed 
under the overlap operation and hence define a topology base $\mathfrak{B}_U$. To obtain a topological base that is 
as rich as the Euclidean base $\mathfrak{B}_E(\mathcal{E})$, the collection of cover regions should be sufficiently 
large (certainly infinite). However, one can generate much more modest bases and topologies using finite covers. In 
particular, one can construct a topology of the environment starting from the place fields covering the environment 
$\mathcal{E}$,
\begin{equation}
\mathcal{E}=\cup_{c=1}^{N_c}\pi_c
\label{pfcover}
\end{equation}
and build a discrete approximation to the Euclidean topology base from the place field domains and their intersection 
closure (Fig.~\ref{Fig3} and Fig.~\ref{Fig8}).

\emph{Example 3: Alexandrov topology on a simplicial complex}. In a simplicial complex $\Sigma$, a neighborhood 
$U_{\sigma}$ of a simplex $\sigma$ is formed by the set of simplexes $\sigma_m$, $m =1,..., n_{\sigma}$, that 
include $\sigma$ (Fig.~\ref{Fig2}A). It can be verified directly that the unions and the intersections of so-defined 
neighborhoods produce another neighborhood from $\Omega(\mathcal{A}({\Sigma}))$, in accordance with the 
Hausdorff axioms \citep{Alexandroff}. 

The overlap of all the neighborhoods containing a given simplex $\sigma$, 
$U_{\sigma} = \cap_m U_{\sigma_m}$, is its minimal neighborhood. The minimal neighborhoods form a topology 
base in finitary space $\mathcal{A}_{\Sigma}$, which defines the Alexandrov topology $\Omega(\mathcal{A}({\Sigma}))$
(Fig.~\ref{Fig8}). In particular, the  Alexandrov topology is defined for all the examples discussed in Section 2: the 
nerve complex $\mathcal{N}$, the temporal complex $\mathcal{T}$ and the cell assembly complex $\mathcal{T}_{CA}$. 

\textbf{Continuous mappings between topological spaces}. A space $X$ maps continuously onto a space $Y$, $f: X \to Y$, 
if each topological neighborhood in $Y$ is an $f$-image of a topological neighborhood in $X$ (for precise discussions see
\citep{Munkres}). If two spaces $X$ and $Y$ map continuously onto each other, then they are topologically equivalent. 
An example of topological equivalence is a continuous deformation of $X$ into $Y$ (one can imagine the corresponding 
deformation of the neighborhoods of $X$ into the neighborhoods of $Y$ that does not violate the mutual overlap, 
containment and adjacency relationships between the neighborhoods). In contrast, if $X$ cannot be transformed into $Y$ 
without adding or removing neighborhoods and points, then $X$ and $Y$ are topologically distinct. 
For example, if a space $Y$ contains an extra hole, then the topology on $Y$ lacks neighborhoods that relate the ``missing'' 
points (contents of the hole) and points outside of the hole. For this reason, a mismatch in the number of holes, handles, 
connectivity components and similar qualitative features serves as immediate indicators of topological inequivalence of spaces. 

It is important to notice, that if the space $X$ has a richer topology (i.e., a larger set of topological neighborhoods) 
than $Y$, then a continuous mapping $f: X \to Y$ may exist, but an inverse mapping, $g: Y \to X$, will not. 
For example, the rich Euclidean topology of the environment $\mathcal{E}$ can map continuously into the finitary 
topology of the memory space $\mathcal{M}$, because many neighborhoods of $\mathcal{E}$ may map into a 
single neighborhood of $\mathcal{M}$. The converse is not true: no mapping can reproduce the infinity of open 
sets in $\mathcal{E}$ from finite set of neighborhoods in $\mathcal{M}$.

\textbf{A continuous mapping of the environment into the memory space} can be constructed as follows. Let us 
consider first the coactivity complex $\mathcal{T}$ and a spatial mapping, $f: \mathcal{M}_{\mathcal{T}} 
\to \mathcal{E}$, that ascribes the Cartesian $(x,y)$ coordinates to the spikes according to the animal's location 
at the time of spiking \citep{Babichev1} (Fig.~\ref{Fig8}G). This function maps the activity of an individual place cell 
into its place field, $f: r_i \to \pi_i$, and the firing pattern of a place cell combination $\sigma$ into its simplex field 
$l_{\sigma}$---the domain where all the cells in $\sigma$ are active, $f: \sigma \to l_{\sigma}$. Notice that simplex 
fields exist for all (not only maximal) simplexes of $\mathcal{T}$. Assuming that some combination of place cells is 
active at every location of the environment (a physiologically justified assumption), implies that the simplex fields 
form a cover of $\mathcal{E}$,
\begin{equation}
\mathcal{E}=\cup_{\sigma=1}^{N_{\sigma}}l_{\sigma}
\label{sxcover}
\end{equation}
Since simplexes of $\mathcal{T}$ may overlap with or include one another, the corresponding simplex fields may 
also overlap. However, for every simplex $\sigma$ there generically exists a subregion of its simplex field---the 
\emph{atomic} region $a_{\sigma}$---where \emph{only} this specific combination of cells is active. The name 
``atomic'' emphasizes that these regions cannot be subdivided any further based on the information provided by 
place cell coactivity (a nonempty overlap of $a_{\sigma}$ with any other region yields $a_{\sigma}$) and that they 
are disjoint ($a_{\sigma} \cap a'_{\sigma} = \emptyset$ for $\sigma \neq\sigma'$). As a result, they form a partition 
of the environment---the atomic decomposition of the cover:
\begin{equation}
\mathcal{E}=\sqcup_{\sigma=1}^{N_{\sigma}}a_{\sigma}
\label{dissxcover}
\end{equation}
which may be viewed as the ultimate discretization of space produced by the given place field map.

Since each atomic element corresponds to a particular simplex $\sigma$ of $\mathcal{T}$, it also defines a point 
$x_{\sigma}$ of $\mathcal{A}_{\mathcal{T}}$, and hence an element of the memory space $\mathcal{M}_{\mathcal{T}}$.
Consider now a reverse mapping, $F: \mathcal{E} \to \mathcal{M}_{\mathcal{T}}$, in which every point 
$r = (x, y)$ of the environment contained in the atomic region $a_{\sigma}$ maps into the corresponding 
point $x_{\sigma}$ of $\mathcal{M}_{\mathcal{T}}$. By construction, every base (minimal) neighborhood in memory 
space $\Omega(\mathcal{M}_{\mathcal{T}})$ is an image of a base neighborhood in the Euclidean topology of the 
environment, $\Omega(\mathcal{E})$, and hence $F$ is continuous map.

\textbf{Continuity in memory space encoded by the cell assembly network}. A similar argument applies to the 
memory space generated by the cell assembly complex $\mathcal{T}_{CA}$. Similarly to the previous case, 
we assume that at least one cell assembly or its subassembly is active in every location of the environment 
\citep{Babichev2} and hence that the place cell (sub)assembly fields $\mathfrak{l}_{\sigma}$ form a cover
\begin{equation}
\mathcal{E}=\cup_{\sigma=1}^{N_{\sigma}}\mathfrak{l}_{\sigma}
\label{cacover}
\end{equation}
The intersection closure of the cell assembly cover yields the decomposition of the environment into the 
non-overlapping atomic regions $\mathfrak{a}_k$, which form a partition of the environment,
\begin{equation}
\mathcal{E}=\sqcup_{k=1}^{N_{k}}\mathfrak{a}_{\sigma}
\label{dissxcover}
\end{equation}
Since every point of the environment belongs to one atomic region that corresponds to a particular minimal neighborhood of the 
memory space, we have a continuous mapping from $\mathcal{E}$ to $\mathcal{T}_{CA}$ and hence $\mathcal{M}_{\mathcal{E}}$.

Alternatively, one can establish continuity of $\mathcal{E}$ to $\mathcal{T}_{CA}$ by constructing a simplicial 
mapping from the coactivity complex $\mathcal{T}$ to its subcomplex $\mathcal{T}_{CA}$, based on the observation 
that both complexes are connected, have finite order, free fundamental groups and identical homologies \citep{Babichev2}. 

\textbf{Stong Matrix}. The numerical analyses of the finite memory spaces were carried out in terms of the Stong 
matrices. If a finite topological space $X$ contains $N$ minimal neighborhoods, $U_1$, $U_2$,..., $U_{N}$, then 
the topological structure on $X$ is uniquely defined by a matrix $M_{ij}$, defined as following:

\begin{enumerate}
\item $M_{ii} =$ number of points that fall inside of the neighborhood $U_i$;
\item if $U_i$ is the immediate neighborhood of $U_j$, $M_{ij} = 1$ and $M_{ji} = -1$;
\item $M_{ij} = 0$ otherwise;
\end{enumerate}

Conversely, every integer matrix satisfying the requirements 1-3 describes a finite topological space $\mathcal{A}$ \citep{Stong}.

For two finitary spaces $\mathcal{A}$ and $\mathcal{B}$, topological equivalence follows from the equivalence of 
the corresponding Stong matrices: $\mathcal{A}$ is equivalent to $\mathcal{B}$, if the topology $\Omega(\mathcal{A})$ 
can be obtained from $\Omega(\mathcal{B})$ by re-indexing the minimal neighborhoods. In other words, $\mathcal{A}$ 
and $\mathcal{B}$ are topologically equivalent if the Stong matrix $M_{\mathcal{A}}$ can be obtained from the Stong 
matrix $M_{\mathcal{B}}$ by a permutation of rows and columns, otherwise they are topologically distinct \citep{Stong}.

\textbf{Reduction of a Stong matrix}. If minimal neighborhood $U_i$ is contained in a single immediate neighborhood 
$U_k$, then it only adds granularity to the Alexandrov space $\mathcal{A}$. Tbe latter can then be coarsened by removing 
$U_i$. If, as a result of coarsening, the neighborhoods separating two points $p_1$ and $p_2$ disappear, then they fuse into a 
single point. This yields a ``reduced'' Alexandrov space $\mathcal{A}'\equiv\mathcal{A}^{{1}}$ that is weakly homotopically 
equivalent to $\mathcal{A}\equiv\mathcal{A}^{{0}}$ \citep{Stong,Osaki}. Such coarsening procedure can be applied multiple 
times: the resulting chain of transformation of $\mathcal{A}$ can be viewed as a discrete homotopy process, $\mathcal{A}^{(0)} \to\mathcal{A}^{(1)}\to\mathcal{A}^{(2)} \to...\to\mathcal{A}^{(n)}$, leading to more and more ``coarse'' topologies 
(Fig.~\ref{Fig3}). 

The numerical procedure implementing the Alexandrov space reduction is as follows. If a column $m_i$ of a Stong matrix 
contains only one non-zero element $m_{ik}$, it is removed along with the corresponding row, then the $n \times n$ matrix 
$M_{\mathcal{A}}$ reduces to a $(n-1) \times (n-1)$ matrix $M'_{\mathcal{A}}$. Eventually, the Stong matrix reduces to 
a ``core'' form that cannot be reduced any further; the corresponding Alexandrov space $\mathcal{C}_{\mathcal{A}}$ is 
referred to as the core of the original Alexandrov space $\mathcal{A}$. The reduction process is illustrated in Suppl. 
Movies~\citep{SupMov4}.

\textbf{Proximity between topologies}. One can quantify difference between finite topologies $\Omega_1$ and $\Omega_2$ by 
estimating the norm of the difference between the corresponding Stong matrices $M_1$ and $M_2$, minimized over the set 
$P$ of all row and column permutations, 
\begin{equation}
D_P(M_1,M_2) = \min_{P}|P(M_1) - M_2|.
\label{minP}
\end{equation}
As a simpler option, one can evaluate the distance between the reduced row echelon forms of the Stong matrices,
\begin{equation}
D(M_1,M_2) = |(\textsf{rref}(M_1)-\textsf{rref}(M_2)|,
\label{rref}
\end{equation}
illustrated Fig.~\ref{Fig4}. Clearly, both distances (\ref{minP}) and (\ref{rref}) vanish if the matrices $M_1$ and $M_2$ are 
equivalent, i.e., if the corresponding memory spaces are homeomorphic. 

\section*{Acknowledgments}
\label{section:acknow}

The work was supported by the NSF 1422438 grant.

\bibliographystyle{frontiersinSCNS_ENG_HUMS}

\newpage
\beginsupplement

\section{Supplementary Figures}

\begin{figure}[H] 
\includegraphics[scale=0.87]{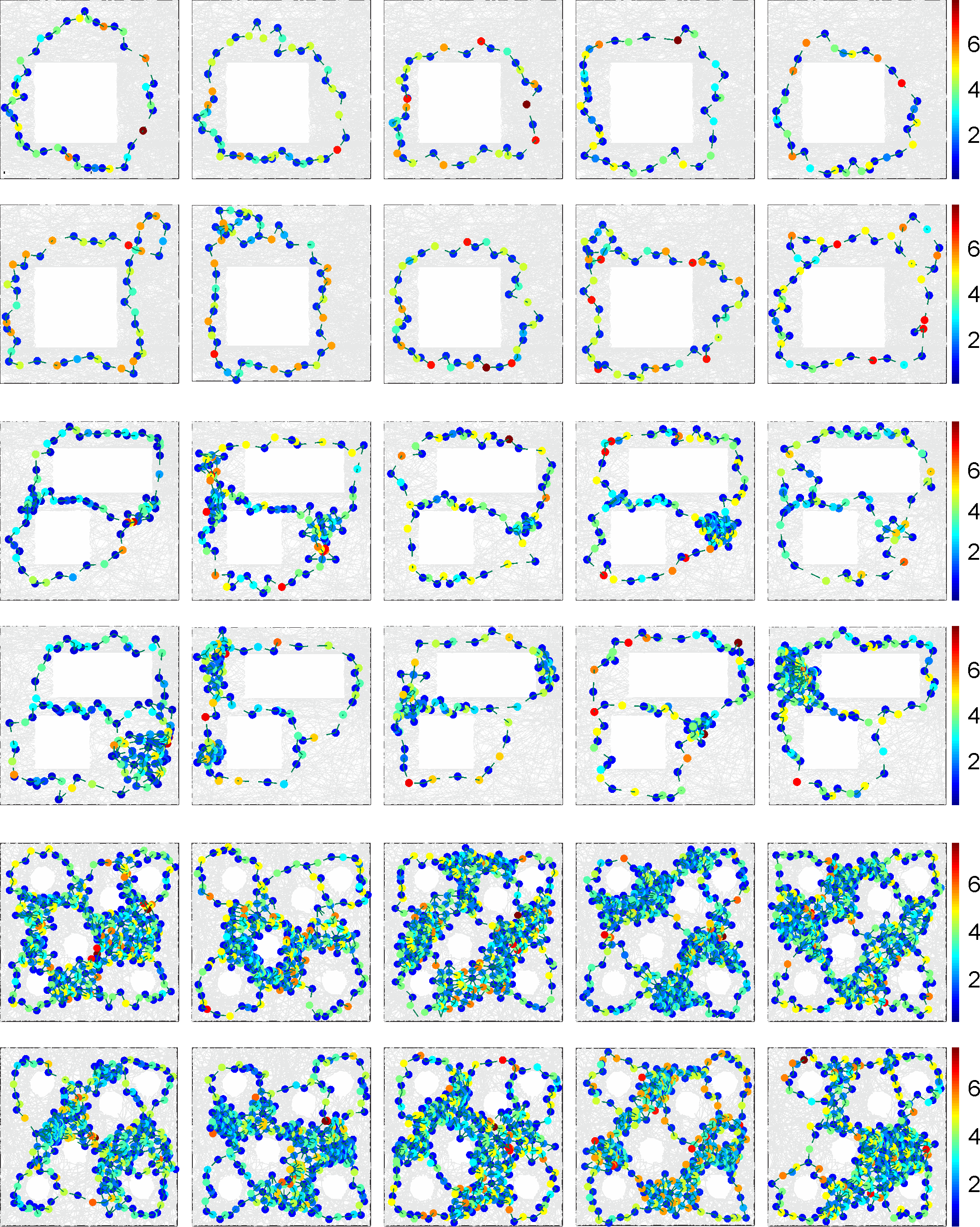}
\caption{{\footnotesize\textbf{Cores of the memory spaces in three environments}. (\textbf{A}) The figure demonstrates cores of the memory 
spaces obtained for ten different place field maps in the three environments shown on Fig.~\ref{Fig3}. The shapes of the cores depend on the
map and on the reduction sequence, however they capture the structure of the environment and approximate its topological core.}}
\label{SFig1}
\end{figure} 

\end{document}